\documentclass[runningheads]{llncs}
\usepackage[T1]{fontenc}
\usepackage{amsmath}
\usepackage{amssymb}
\usepackage{xcolor}
\usepackage[hidelinks]{hyperref}
\usepackage{cite}
\usepackage{orcidlink}
\usepackage{graphicx}
\usepackage{algorithm}
\usepackage{algorithmic}
\usepackage{tikz}

\begin{document}
\title{Approximate Optimal Active Learning of Decision Trees}
%
%
%

\author{Zunchen Huang \orcidlink{0000-0002-5837-9960} \and Chenglu Jin \orcidlink{0000-0001-6306-8019}}
%
%
\institute{Centrum Wiskunde \& Informatica, Science Park 123, Amsterdam, The Netherlands \email{\{zunchen.huang, chenglu.jin\}@cwi.nl}}

\maketitle              

\begin{abstract}
We consider the problem of actively learning an unknown binary decision tree using only membership queries, a setting in which the learner must reason about a large hypothesis space while maintaining formal guarantees.  Rather than enumerating candidate trees or relying on heuristic impurity or entropy measures, we encode the entire space of bounded-depth decision trees symbolically in SAT formulas.  We propose a symbolic method for active learning of decision trees, in which approximate model counting is used to estimate the reduction of the hypothesis space caused by each potential query, enabling near-optimal query selection without full model enumeration. The resulting learner incrementally strengthens a CNF representation based on observed query outcomes, and approximate model counter ApproxMC is invoked to quantify the remaining version space in a sound and scalable manner. Additionally, when ApproxMC stagnates, a functional equivalence check is performed to verify that all remaining hypotheses are functionally identical. Experiments on decision trees show that the method reliably converges to the correct model using only a handful of queries, while retaining a rigorous SAT-based foundation suitable for formal analysis and verification.

\end{abstract}
\section{Introduction}
\label{sec:intro}

Decision trees are a widely used model for representing Boolean functions across multiple domains, including program analysis~\cite{DBLP:conf/birthday/CousotCM10, DBLP:conf/sas/UrbanM14, DBLP:conf/sas/ChenC15, DBLP:conf/issre/FrancisLMP04}, verification~\cite{DBLP:journals/corr/KrishnaPW15, DBLP:conf/popl/0001NMR16}, and machine learning~\cite{DBLP:books/wa/BreimanFOS84, DBLP:journals/ml/Quinlan86, DBLP:journals/tcad/HuangJ23}. In formal methods scenarios, decision trees naturally arise when modeling temporal logic, classification decisions, or conditional behaviors in programs~\cite{DBLP:conf/hybrid/BombaraVPYB16, DBLP:conf/jelia/BrunelloSS19, DBLP:journals/corr/abs-2105-11508, DBLP:journals/tcps/BombaraB21, DBLP:conf/l4dc/AasiCVB23, DBLP:conf/cdc/LiangCKV24}. Unlike deep neural network with complex, layered architectures and difficulty in interpreting how predictions are made, decision trees are interpretable, compact, and can represent arbitrary Boolean functions of bounded complexity, making them suitable for symbolic reasoning on various systems.

In many practical applications, we are confronted with a fundamental problem: given black-box access to a Boolean function implemented as a decision tree, can we efficiently reconstruct the underlying function with as few queries as possible? Black-box access is typically achieved through a \emph{membership oracle}, which evaluates the function on input provided by the learner. This setup is relevant in software testing~\cite{DBLP:conf/issta/BowringRH04, DBLP:journals/tse/XieN05, DBLP:conf/icse/MarianiPRS11}, formal verification~\cite{DBLP:conf/forte/PeledVY99, DBLP:conf/cav/SenVA04, DBLP:conf/rv/AichernigT17, DBLP:conf/rv/ShijuboWS21}, and security analyses~\cite{DBLP:journals/ieeesp/McGrawP04, DBLP:conf/sp/BauBGM10, DBLP:conf/acsac/LiX11, DBLP:conf/ndss/PellegrinoB14}, where direct inspection of the implementation may be impossible, expensive, unsafe, or breaking privacy commitment. However, controlled and highly selective queries can be made to infer systems behavior and useful properties efficiently.

Traditional approaches to active learning of Boolean functions fall into two main categories. First, \emph{enumeration-based methods} attempt to explicitly maintain and prune the entire hypothesis space of candidate functions consistent with the observed queries~\cite{DBLP:books/cu/10/SloanSTCH10, DBLP:conf/cav/ChenW12}. While these methods are conceptually straightforward, they scale poorly: the number of candidate decision trees grows exponentially with the number of features and the tree depth, quickly becoming intractable even for moderate problem sizes. Second, \emph{heuristic-based methods} select queries based on information gain, or other statistical criteria~\cite{DBLP:conf/nips/BalcanF13}. Although these heuristics can guide query selection without explicit enumeration, they provide no formal guarantees on the reduction of the hypothesis space and may require many redundant queries, especially when the underlying function exhibits intricate dependencies between features.

\begin{figure}
	\centering
	\includegraphics[width=1\textwidth]{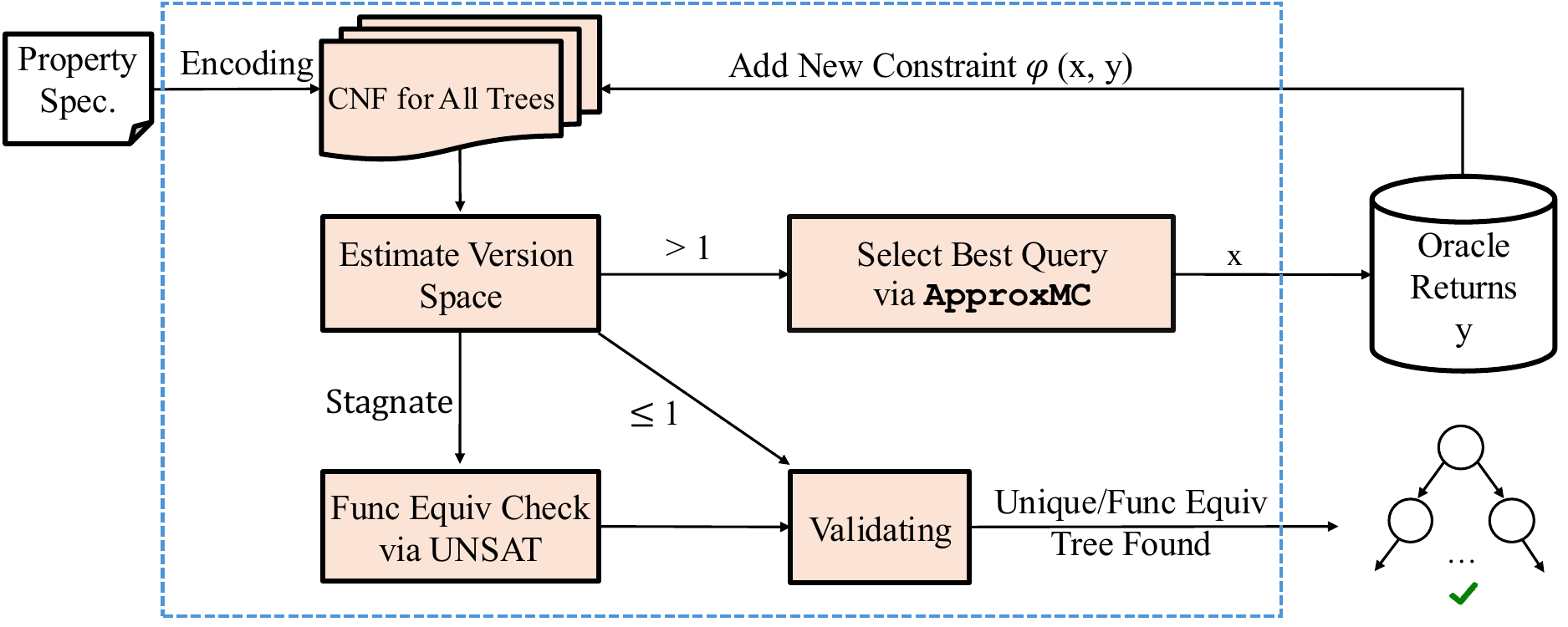}
	\caption{\textsc{SAMCAL}: Approx. Model Counting Guided Symbolic Active Learning Method}
	\label{fig:flow}
\end{figure}

In this work, we introduce a symbolic, model counting guided approach that bridges the gap between formal methods and active learning, which is illustrated in Figure~\ref{fig:flow}. The symbolic reasoning engine is within the blue dotted box. Our key insight is that the space of all decision trees of a given depth and feature set can be encoded compactly as a \emph{propositional formula} in conjunctive normal form (CNF) from a user provided property specification, i.e., decision tree topology. Each model of the CNF corresponds to a valid decision tree, satisfying constraints on internal node feature selections and leaf output assignments. Membership queries are then used to prune this hypothesis space: every observed input-output pair imposes additional CNF constraints, eliminating all candidate trees inconsistent with the observation.

To efficiently estimate the remaining hypothesis space after each query, we leverage \emph{approximate model counting}, using the state-of-the-art \texttt{ApproxMC}~\cite{DBLP:conf/cp/ChakrabortyMV13} algorithm. \texttt{ApproxMC} provides a scalable method for estimating the number of satisfying assignments of a CNF formula without exhaustively enumerating all models. By using the counts to estimate the effect of potential queries on the version space, our learner selects the input that approximately halves the number of consistent hypotheses. It is possible that \texttt{ApproxMC} is unable to reduce the hypothesis space at the final iteration rounds. The functional equivalence check is designed to verify whether two remaining hypotheses, e.g., two decision trees or logical models, are functionally identical. This is achieved by encoding that the outputs of the two models should differ at least on one input. A SAT solver is invoked to determine if this assumption is unsatisfiable. This module is crucial to ensuring the correctness of the learning process, where we can confidently collapse equivalent hypotheses into one and continue refining and terminate the learning process.

By combining symbolic SAT-based reasoning with approximate model counting, our approach provides a rigorous framework for active learning in formal methods settings. Unlike purely heuristic methods, our technique maintains a formal representation of the version space, enabling guarantees about the reduction of hypotheses after each query. At the same time, approximate counting ensures scalability, avoiding the need for full enumeration of candidate decision trees. Consequently, this method represents a promising approach for tasks such as automated program analysis, testing of black-box components, and verification of systems with hidden components, where query efficiency and formal guarantees are both critical.

The main contributions of this work are as follows:

\begin{itemize}
    \item \textbf{Formal encoding of decision tree hypothesis spaces.} We develop a systematic method for encoding all depth-$d$ decision trees over $n$ binary features. The resulting CNF is suitable for symbolic reasoning and approximate model counting. 
    
    \item \textbf{Symbolic active learning framework.} We demonstrate how approximate model counting can guide query selection. Each candidate query is evaluated for its expected reduction in the hypothesis space, and the query maximizing this reduction is selected. This allows us to actively prune large hypothesis space efficiently.
    
    \item \textbf{Theoretical and practical analysis.} We show that each selected query approximately halves the hypothesis space, both theoretically and empirically. This near-halving property provides formal insight into why the method scales better than naive enumeration.

    \item \textbf{Evaluation on unknown decision trees.} We implement a Python tool, and it successfully reconstructs the hidden function, where the largest initial hypothesis space is over $10^{15}$, using only a small number of membership queries, illustrating the feasibility and showing the scalability of our method.

\end{itemize}

The remainder of the paper is outlined as follows. In Section~\ref{sec:form}, we formalize the problem and analyze the version space with approximate model counting. In Section~\ref{sec:algo}, we present the algorithm for approximate optimal active learning. In Section~\ref{sec:encoding}, we define the general SAT encoding . In Sections~\ref{sec:motivation} and~\ref{sec:eval}, we show the result on a motivation example and more general settings on the trees. We discuss the scalability, limitations, and related works in Sections~\ref{sec:dis} and~\ref{sec:rel}. We finally present the conclusion in Section~\ref{sec:conc}.

\section{Problem Formulation}
\label{sec:form}

We formalize the active learning problem for decision trees. Let $n$ denote the number of Boolean features, and $d$ the depth of the decision tree. 

\subsection{Hypothesis Space}
Let $H$ denote the set of all decision trees of depth $d$ over $n$ Boolean features. Each hypothesis $h \in H$ is fully defined by:
\begin{itemize}
    \item \textbf{Internal node assignments:} For each internal node $i$, a feature $T_i \in \{1, \dots, n\}$ is selected. 

    \item \textbf{Leaf outputs:} Each leaf $\ell$ is assigned a Boolean output $B_\ell \in \{0,1\}$.
\end{itemize}

\subsection{Version Space}
Given a set of queries and observed oracle responses up to step $t$, the \emph{version space} $V_t$ is the subset of hypotheses consistent with all observations:
\[
V_t = \{ h \in H \mid h(x) = y, \forall (x,y) \text{ queried up to step } t\}.
\]

\subsection{Membership Oracle}
We assume access to a black-box \emph{membership oracle}:
\[
\text{oracle}: \{0,1\}^n \to \{0,1\}, \quad x \mapsto f(x),
\]
which returns the output of the unknown target function for any input $x$. The objective of active learning is to select a sequence of queries that efficiently shrinks $V_t$ until $|V_t| = 1$, i.e., the target hypothesis is uniquely identified.

\subsection{Approximate Halving Analysis}
\label{sec:half}

For any (remaining) candidate input $x$, let
\[
V_t^0 = \{ h \in V_t \mid h(x) = 0 \}, \quad
V_t^1 = \{ h \in V_t \mid h(x) = 1 \}.
\]
The ideal query maximizes the minimum reduction in the version space:
\[
x^* = \arg\max_{x \in \{0,1\}^n} \min(|V_t^0|, |V_t^1|).
\]

\subsection{Approximate Counting with \texttt{ApproxMC}}
Exact computation of $|V_t^0|$ and $|V_t^1|$ is generally infeasible. We leverage \emph{approximate model counting} (\texttt{ApproxMC}) to estimate these values:
\[
\tilde{c}_0 = \text{ApproxMC}(V_t \wedge (h(x) = 0)), \quad
\tilde{c}_1 = \text{ApproxMC}(V_t \wedge (h(x) = 1)).
\]
We then select
\[
x^* = \arg\max_x \min(\tilde{c}_0, \tilde{c}_1),
\]
which approximately halves the version space at each step.

\subsection{Geometric Decay of Version Space}
Let $V_{t+1}$ denote the version space after observing the oracle output for $x^*$. Then
\[
|V_{t+1}| \approx \frac{1}{2} |V_t|,
\]
assuming the query is near-optimal. Iteratively applying this strategy yields logarithmic query complexity in the size of the initial hypothesis space $|\mathcal{H}|$:
\[
\text{Number of queries} \lesssim \log_2 |\mathcal{H}|.
\]

This forms the theoretical foundation for our approximate optimal active learning framework for decision trees.

\section{Approximate Optimal Active Learning Algorithm}
\label{sec:algo}

We now present the algorithm for approximately optimal active learning of decision trees using model counting guided queries. The algorithm iteratively selects inputs that approximately halve the remaining version space, based on estimates from \texttt{ApproxMC}.  

\subsection{Algorithm Overview}

\begin{algorithm}[htbp]
\caption{High Level Procedure of the Proposed Method \textsc{SAMCAL}}
\label{algo:high}
\begin{algorithmic}[1]
\STATE \textbf{Input:} Tree spec, ApproxMC, oracle
\STATE \( \phi \gets \text{EncodeCNFTrees()} \)
\STATE \( V_t \gets \{\phi\} \)  \COMMENT{--- Initial version space ---}
\WHILE{\( |V_t| > 1 \)}
    \STATE \( x^* \gets \text{ApproxMC}(V_t) \)  \COMMENT{--- Select optimal query ---}
    \STATE \( y^* \gets \text{oracle}(x^*) \)
    \STATE Add constraint \( \phi_{x^*, y^*} \) to \( V_t \)
    \STATE \( V_{t+1} \gets \{ h \in V_t \mid h(x^*) = y^* \} \)
\ENDWHILE
\STATE \textbf{Output:} Tree \( h^* \in V_t \)
\end{algorithmic}
\end{algorithm}

We show the high-level procedure of our method in Algorithm~\ref{algo:high}. Let $\mathcal{H}$ be the hypothesis space of all decision trees of a given depth and feature set, and $V_t \subseteq \mathcal{H}$ the version space after $t$ queries. At each step, the algorithm performs the following operations: For each unqueried input $x \in \{0,1\}^n$, use \texttt{ApproxMC} to estimate the number of hypotheses consistent with $V_t$ that would output 0 or 1. The score of input $x$ is defined as $\min(\tilde{c}_0, \tilde{c}_1)$. Choose the input $x^*$ with the highest score, which approximately halves the version space. Obtain the output $y = \text{oracle}(x^*)$. Restrict $V_t$ to hypotheses consistent with the new observation: $V_{t+1} = \{ h \in V_t \mid h(x^*) = y \}.$ Continue until the version space is reduced to a single hypothesis, i.e., $|V_t| = 1$, or a stagnation is found and all remaining trees are checked whether they are functional equivalent.

\subsection{Algorithm Pseudocode}
	
\begin{algorithm}[htbp]
    \caption{Approximate Optimal Active Learning}
    \label{algo:func}
    \begin{algorithmic}[1]
        \STATE $\mathcal{H}$ = calculate\_H(TreeSpec)
        \STATE Initialize version-space CNF: $V \gets \mathcal{H}$   \hfill (initial tree hypothesis space)
        \STATE Initialize queried set: Queried $ \gets \emptyset$
        \STATE $\tilde{C}_{\mathrm{prev}} \gets \text{ApproxMC}(V)$
        \STATE round $\gets 0$
        \WHILE{true}
        \FORALL{$x \in \{0,1\}^n \setminus \text{Queried}$}
        \STATE $\tilde{c}_0 \gets \text{ApproxMC}(V \wedge (h(x) = 0))$
        \STATE $\tilde{c}_1 \gets \text{ApproxMC}(V \wedge (h(x) = 1))$
        \STATE $score(x) \gets \min(\tilde{c}_0, \tilde{c}_1)$
        \ENDFOR
        \STATE $x^\ast \gets \arg\max_x score(x)$
        \STATE $y^\ast \gets \text{oracle}(x^\ast)$
        \STATE $\text{Queried} \gets \text{Queried} \cup \{x^\ast\}$
        \STATE $V \gets V \wedge (V_{x^\ast} = y^\ast)$
        \STATE $\tilde{C}_{\mathrm{new}} \gets \text{ApproxMC}(V)$
        
        \COMMENT{--- Unique Hypothesis Found ---}
        \IF{$\tilde{C}_{\mathrm{new}} \le 1$}
        \STATE \textbf{return} $V$
        \ENDIF
        
        \COMMENT{--- Check for ApproxMC stagnation ---}
        \IF{$|\tilde{C}_{\mathrm{new}} - \tilde{C}_{\mathrm{prev}}| < 1$}
        \STATE Construct two copies $V^{(1)}, V^{(2)}$ of $V$
        \STATE Let $G \gets V^{(1)} \land V^{(2)} \land \exists x\ (f_1(x) \neq f_2(x))$
        \IF{$\text{Solve}(G) = \text{UNSAT}$}
        \STATE \textbf{return} Any model of $V$ \hfill (all remaining trees are functionally identical)
        \ENDIF
        \ENDIF
        \STATE $\tilde{C}_{\mathrm{prev}} \gets \tilde{C}_{\mathrm{new}}$
        \STATE round $\gets$ round + 1
        \IF{round $\ge$ MAX\_ROUND}
        \STATE \textbf{return} "NO UNIQUE TREE FOUND"
        \ENDIF
        \ENDWHILE
    \end{algorithmic}
\end{algorithm}

In Algorithm~\ref{algo:func}, we show our algorithm in detail. We first calculate $\mathcal{H}$ from the tree specification, i.e., depth and feature numbers of tree topology. To make it general, we assume the tree is a full binary tree. In lines 2-3, we encode the unknown tree into a CNF formula and initialize a set Queried to store all the inputs that have been asked to the oracle. In line 4, we invoke the model counter to approximate the models for the first time. Then initialize the round variable. In the while-loop from line 6, we select the input that can approximately obtain the highest score and store it to $x^*$ in line 12, which can maximize the min reduction in the remaining version space. Then we query the oracle and receive $y^*$ in line 13. We add the queried input into the Queried set in line 14. And add the corresponding constraint $\phi(x^*, y^*)$ into the CNF formula and updated $V$ in line 15. In line 16, we invoke the model counter again to approximate the size of the updated $V$. If $\tilde{C}_{\mathrm{new}}$ is $\le 1$, then we return the unique hypothesis. Otherwise, we continue to check if the model counter stops making progress by checking the old and new $V$ sizes difference smaller than 1 in line 20. If stagnation is detected, we verify that if there exists an input $x$ and two different hypotheses would have different outputs on $x$, and if this condition is checked with output UNSAT in line 23, then we conclude functional equivalence for all remaining trees and return any of the hypotheses. To guarantee that our algorithm always terminates, we add a sanity check in lines 28-31 that if the maximum round is reached, then we fail to find the unique hypothesis, where the maximum round is determined by the feature number. However, we never encounter this scenario in our experiments.

\section{General CNF Encoding for Depth-$d$ Decision Trees}
\label{sec:encoding}

We now present a general propositional encoding for the hypothesis space
$\mathcal{H}_{d,n}$ of decision trees of maximum depth $d$ over $n$ binary
features $x_1,\dots,x_n \in \{0,1\}$.  
This encoding supports exact SAT solving and approximate model counting with \texttt{ApproxMC}, enabling our active learning algorithm to prune the hypothesis space incrementally as oracle responses are received.

A full binary tree of depth $d$ has:
\[
N_{\text{int}} = 2^d - 1 \quad \text{internal nodes}, \qquad
N_{\text{leaf}} = 2^d \quad \text{leaves}.
\]
\begin{lemma}
\label{lem}
Let $\mathcal{H}_{d,n}$ be the class of full (complete) binary decision trees of depth $d$ over $n$ Boolean features, where
\begin{itemize}
  \item each internal node selects \emph{one} feature from the $n$ features (selection may repeat across nodes),
  \item each leaf stores a Boolean output in $\{0,1\}$.
\end{itemize}
Then the number of distinct hypotheses in $\mathcal{H}_{d,n}$ is
\[
|\mathcal{H}_{d,n}| \;=\; n^{\,2^d-1}\; \cdot\; 2^{\,2^d}.
\]
\end{lemma}

The proof is shown in the Appendix.

\subsection{Internal Node Feature Selection}
\label{sec:internal-selection}

Each internal node $u$ selects exactly one feature to test.  
We introduce variables $S_{u,i} \quad (u = 1\dots N_{\text{int}},\; i = 1\dots n) $
representing node $u$ tests feature $x_i$.

\paragraph{Constraint \(\phi_{\text{sel}}\).} At least one feature must be chosen, and no two features may be selected simultaneously. 

\[
\phi_{\text{sel}}
\;=\;
\vee_{i=1}^n S_{u,i} 
\bigwedge
\wedge_{1 \le i < j \le n}
(\lnot S_{u,i} \,\lor\, \lnot S_{u,j}).
\]

\subsection{Leaf Labeling}
\label{sec:labeling}

Each leaf $v$ has exactly one class from the label set $\mathcal{Y}$.
We introduce variables: $ L_{v,c}, c \in \mathcal{Y}$.

\paragraph{Constraint \(\phi_{\text{leaf}}\).} Similarly as internal node featue selection, 
\[
\phi_{\text{sel}}
\;=\;
\vee_{c \in \mathcal{Y}} L_{v,c}
\bigwedge
\wedge_{c < c'}
(\lnot L_{v,c} \lor \lnot L_{v,c'}).
\]

\subsection{Reachability of Nodes}
\label{sec:reachability}

For each input $x$ in the set of evaluated examples, we introduce
variables: $R_u(x)$, which means input $x$ can reach node $u$.

\paragraph{Constraint \(\phi_{\text{root}}\).}
The root is always reachable: $\phi_{\text{root}} = R_{r}(x).$

\paragraph{Propagation Constraints \(\phi_{\text{prop}}\).}

Suppose internal node $u$ tests feature $x_i$.  
Then:

\[
R_{\text{left}}
\iff
\big(R_u(x) \land x_i\big),
\quad
R_{\text{right}}
\iff
\big(R_u(x) \land \lnot x_i\big).
\]

Since the selected feature is itself encoded by $\{S_{u,i}\}_i$, we write:

\[
\phi_{\text{prop-left}} \;=\;
(\lnot S_{u,i} 
\lor \lnot R_u(x)
\lor \lnot x_i
\lor R_{\text{left}(u)}(x)),
\]

\[
\phi_{\text{prop-right}} \;=\;
(\lnot S_{u,i} 
\lor \lnot R_u(x)
\lor x_i 
\lor R_{\text{right}(u)}(x)).
\]

Full propagation constraint at node $u$:
\[
\phi_{\text{prop}}
=
\bigwedge_{i=1}^n
\big(
\phi_{\text{prop-left}_i}
\;\land\;
\phi_{\text{prop-right}_i}
\big).
\]

\subsection{Oracle Consistency Constraints}
\label{sec:oracle}

Whenever the membership oracle returns $(x^*, y^*)$, every valid hypothesis
must send $x^*$ to a leaf that outputs $y^*$.

For each leaf $v$:

\[
\phi_{\text{oracle}}
=
(\lnot R_v(x^*) \lor L_{v,y^*}).
\]

This constraint is added incrementally after each query, shrinking the
remaining hypothesis space.

\subsection{Putting Everything Together}
\label{sec:phi-all}

The complete CNF encoding is:
\[
\phi_{\text{all}} = \phi_{\text{select}} 
\;\land\; 
\phi_{\text{leaf}} 
\;\land\;
\phi_{\text{root}}
\;\land\;
\phi_{\text{prop}}
\;\land\;
 \phi_{\text{oracle}}
\]

This single SAT instance represents the current space of hypotheses
consistent with all oracle responses observed so far.  
\texttt{ApproxMC} is run over $\phi_{\text{all}}$ to estimate the remaining number
of valid decision trees at each iteration of active learning.

\section{Motivation Example}
\label{sec:motivation}

We now present a concrete worked example illustrating how our \textsc{SAMCAL} active learner incrementally eliminates incorrect decision trees until a unique hypothesis is identified.  We restrict the hypothesis space to:

\begin{itemize}
\item a full binary decision tree of depth $d=2$,
\item $n=3$ binary input features $x_1,x_2,x_3$,
\item binary classification labels $\mathcal{Y}=\{0,1\}$.
\end{itemize}

A depth-$2$ tree has $ N_{\text{int}} = 2^2-1 =3 $ internal nodes, $N_{\text{leaf}} = 2^2 = 4 $ leaves. Each internal node must choose exactly one of the three input features, and each leaf must select either output label 0 or 1.  
Thus, the theory permits: $|\mathcal{H}| = n^{N_{\text{int}}} \cdot 2^{N_{\text{leaf}}}= 3^3 \cdot 2^4 = 432$ candidate trees before any evidence is observed.

\subsection{True Hidden Tree}

For the running example, the oracle corresponds to a fixed hidden
decision tree unknown to the learner:

\[
Tree: 
\begin{cases}
\text{Node 1 tests } x_3,\\
\text{Node 2 tests } x_1,\\
\text{Node 3 tests } x_2,
\end{cases}
\quad
\text{Leaf outputs} = [1,0,1,0].
\]

The learner has no knowledge of this structure except through queries.

\subsection{Incremental Learning Process}

Before any queries, \texttt{ApproxMC} reports: $|\mathcal{H}| \approx 432$. Each round, the learner greedily select the input maximizing the expected reduction in the remaining models and then ask the oracle for output. Table~\ref{tab:approx-results} shows the actual execution from our tool. The first column shows the number of step and indicates the iterative round. The second and third columns show the selected input $x^*$ and $y^*$ from oracle. The last column shows how exactly the hypothesis space has reduced. The hypothesis steadily reduces from step one to step six, achieving near halving reduction. After seven queries, \texttt{ApproxMC} reports one remaining model, indicating that the learner has effectively identified a unique decision tree consistent with all observations without the need of checking functional equivalence. In this case, \texttt{ApproxMC} provides an exact and correct solution. We also notice that input (1,1,0) is not necessary to be queried to find this specific tree, indicating inputs selection is sensitive to ground truth during the learning process.

\begin{table}[h]
\centering
\begin{tabular}{c|c|c|c}
\textbf{Step} & \textbf{Selected Queried Input $x^*$} & \textbf{Oracle Output $y^*$} & \textbf{Approx.\ Remaining} \\
\hline
0 & -- & -- & 432 \\
1 & (0,0,0) & 0 & 216 \\
2 & (1,1,1) & 1 & 108 \\
3 & (0,0,1) & 0 & 72  \\
4 & (0,1,1) & 0 & 35  \\
5 & (1,0,0) & 0 & 17  \\
6 & (1,0,1) & 1 & 9   \\
7 & (0,1,0) & 1 & 1 \\
\end{tabular}
\caption{Approximate model counts over successive active-learning rounds.}
\label{tab:approx-results}
\end{table}

\section{Experiments}
\label{sec:eval}

We have implemented our method in a tool named \texttt{SAMCAL} written in Python with 2k lines of code, which builds on the \textsc{ApproxMC} 4.1.24 approximate model counter~\cite{approxmc-4.1.24} and \textsc{MiniSat} 2.2 SAT solver~\cite{minisat-2.2} within \textsc{PySAT}~\cite{pysatcode}. The tool constructs a CNF encoding for the decision tree hypothesis space, with constraints for internal nodes, leaf outputs, and feature assignments. \textsc{ApproxMC} is used to estimate the number of remaining hypotheses after each query, guiding the active learning process by selecting queries that maximize hypothesis space reduction. When ApproxMC stagnates, \texttt{SAMCAL} invokes \textsc{MiniSat} to check for functional equivalence by verifying whether all remaining hypotheses compute the same Boolean function. If a functional collapse is detected (i.e., all hypotheses are functionally equivalent), the learning process terminates early, ensuring efficiency.

We evaluate our tool under the decision trees setting with depth = [2, 3, 4] and features = [3, 4, 5]. Judging from the parameters, it seems the tree is small, however, when the depth is 4 with 5 features, the hypothesis space is $2 * 10^{15}$ (initial hypothesis size is shown in the last column in Table~\ref{tab:cnf-size}, growing at a speed more than exponential), and which is sufficient to demonstrate our tool's capacity. For each tree setting, we randomly generate one ground truth table consisting of node and feature correspondence and leaf boolean values. Our experiments were designed to answer the following questions:

\begin{itemize}
    \item \textbf{RQ1}: Is our method effectively reduce the hypothesis space to identify the target decision tree?
    
    \item \textbf{RQ2}: How does our method scale when features and depth increase?
    
    \item \textbf{RQ3}: When stagnation occurs on \textsc{ApproxMC}'s counting, does functional equivalence collapse occur?
\end{itemize}

The experiments were conducted on a computer with AMD Ryzen 5 5600X CPU and 32GB memory, running Ubuntu 20.04. In Table~\ref{tab:cnf-size}, we show the initialization of all trees' variables and clauses size in CNF formulas. Columns 1 and 2 shows the feature and depth numbers. The third and fourth column shows the variable and clauses size. We notice that when the feature number is the same, the CNF variables are doubling and clauses sizes are tripling as depth increases. When the depth number is the same, the CNF variables and clauses sizes are doubling as features increases. The depth is more dominant on clauses size than feature numbers. 
	
\begin{table}[htbp]
    \centering
    \begin{tabular}{|c|c|c|c|c|}
        \hline
        \textbf{Features} & \textbf{Depth} & \textbf{CNF Vars} & \textbf{CNF Clauses} & \textbf{Hypothesis Size $|\mathcal{H}|$}\\
        \hline
        3 & 2 & 77 & 324 & 432\\ \hline
        3 & 3 & 157 & 868 & 559,872\\ \hline
        3 & 4 & 317 & 2,404 & 940,369,969,152\\ \hline
        4 & 2 & 144 & 693 & 1,024\\ \hline
        4 & 3 & 292 & 1,841 & 4,194,304\\ \hline
        4 & 4 & 588 & 5,033 & 70,368,744,177,664\\ \hline
        5 & 2 & 275 & 1,473 & 2,000\\ \hline
        5 & 3 & 555 & 3,885 & 20,000,000\\ \hline
        5 & 4 & 1,115 & 10,501 & 2,000,000,000,000,000\\
        \hline
    \end{tabular}
    \caption{Initial CNF Variables and Clauses Size on all Trees}
    \label{tab:cnf-size}
\end{table}

\subsection{Effectiveness - Results for Answering RQ1}
\label{sec:effective}

\begin{figure}
	\centering
	\includegraphics[width=1\textwidth]{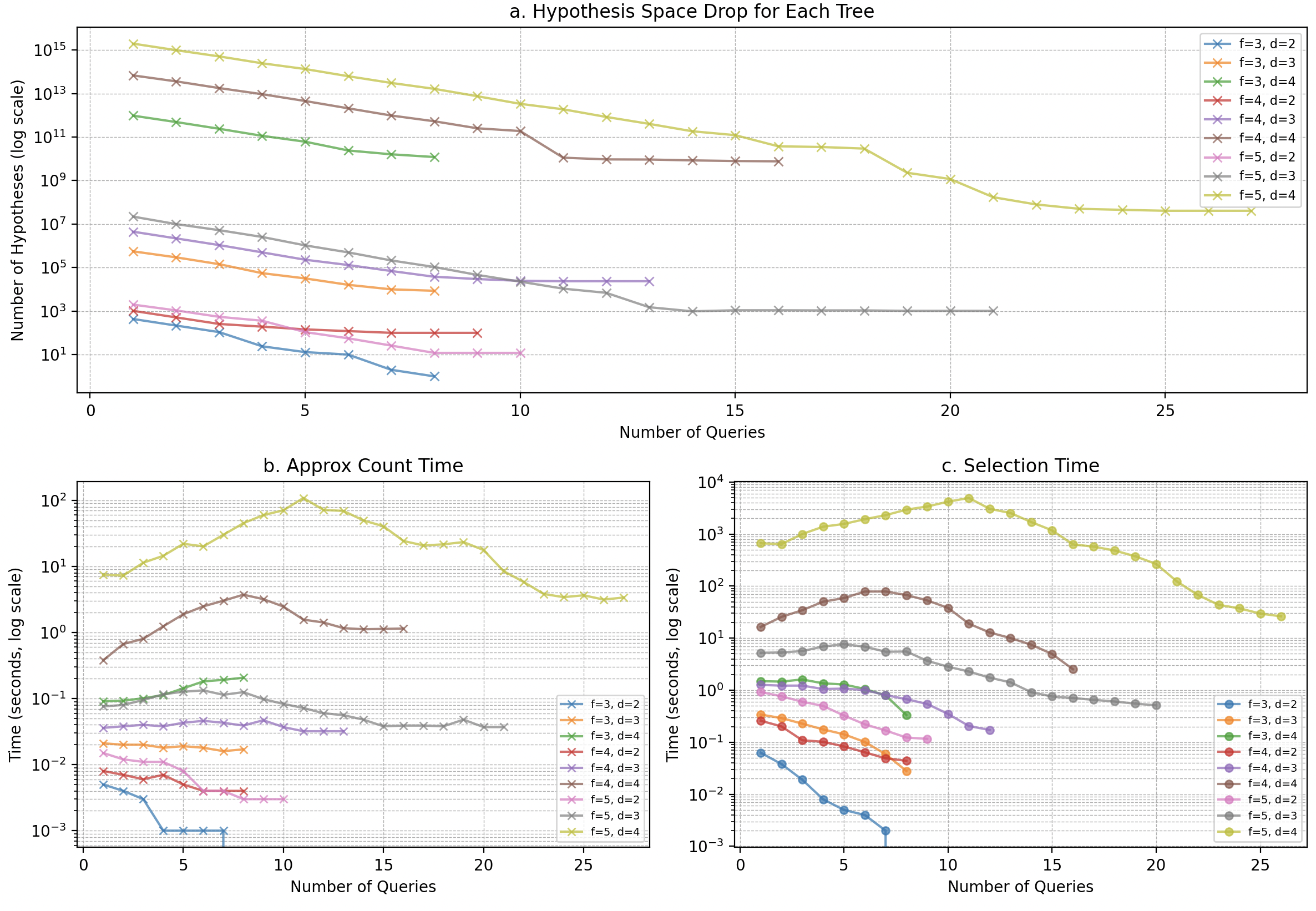}
	\caption{Hypothesis Space Drop, Count, and Selection Time Trend}
	\label{fig:fig1}
\end{figure}

To answer whether our method is effective to reduce the hypothesis and identify the target decision tree. We show in the hypothesis space drop trend for each tree in Figure~\ref{fig:fig1}a. The x-axis denotes the number of queries, and the y-axis denotes the size of hypothesis spaces on a log scale. Across all evaluated configurations (feature counts 3–4 and depths 2–4), our method consistently reduced the hypothesis space at an exponential rate before query 10. For deeper trees such as n=4, d=4 and n=5, d=4, the initial hypothesis spaces were enormous, ranging from hundreds of thousands to several billion, yet the algorithm still achieved rapid reduction, typically stabilizing or collapsing within 8–12 queries. The deepest cases (d=4) started from exceedingly large spaces (up to $10^{15}$), but still exhibited smooth monotone decay under our query strategy. Note in several cases, there is a sudden reduction benefiting from one query, e.g., in query 11 from tree n=4, d=4, there is a sudden space reduction over several magnitude. The reason is that this query is highly informative and remedies the non-optimality of previous queries. In four cases n=4/5, d=3/4), we can see a short flat curve, which indicates approximation of the models making little progress. It still makes progress; otherwise, the curve would terminate earlier. However, this means effectiveness of our method and this behavior will be addressed in Section~\ref{sec:stag}. Overall, these results demonstrate that our selection strategy is highly effective at driving fast version-space collapse regardless of tree depth or feature count.

\subsection{Scalability - Results for Answering RQ2}
\label{sec:scalable}

We show our timing results in Figure~\ref{fig:fig1}b and~\ref{fig:fig1}c. The x-axis denotes the number of queries, and the y-axis denotes the time spent on a log scale. Figure~\ref{fig:fig1}b shows the initial and following approximation of the number of remaining models after adding input-output pair constraint from each query. The cost of querying scales with both the feature count and tree depth, exhibiting predictable growth as the hypothesis space becomes larger. For small and moderate configurations (e.g., n=3/4, d=2/3), both the \textsc{ApproxMC} counting time and the query-selection time remain well under a second. As depth increases to d=4, the per-query cost grows substantially, which is as expected from the exponential hypothesis space, yet remains well-behaved: for f=4 the peak selection time reaches only a few seconds, and for n=5 the most demanding configuration (n=5, d=4) shows a rapid rise to the worst-case of 4k seconds per selection round. Importantly, even in this extreme setting the version space reduces steadily, causing subsequent \textsc{ApproxMC} and selection times to decrease quickly. This behavior indicates that although our method naturally inherits exponential scaling from the underlying combinatorial complexity, the active-learning process benefits from fast hypothesis elimination, keeping the overall runtime manageable and demonstrating practical scalability across all evaluated tree sizes.

\subsection{Functional Equivalence Follows Stagnation - Results for Answering RQ3}
\label{sec:stag}

\begin{table}[htbp]
    \centering
    \begin{tabular}{|c|c|c|c|c|c|c|c|}
        \hline
        \textbf{Tree} & \textbf{Step} & \textbf{\# Before} & \textbf{\# After} & \textbf{Stagnate} & \textbf{Func Equi} & \textbf{Solve Time} & \textbf{Reduction}\\
        \hline
        n4\_d2 & 7$\rightarrow$8 & 100 & 1 & yes & yes & < 1s & 50\% (8)\\ 
        \hline
        n4\_d3 & 11$\rightarrow$12 & 23,552 & 1 & yes & yes & < 1s & 25\% (5)\\
        \hline
        n5\_d2 & 8$\rightarrow$9 & 12 & 1 & yes & yes & < 1s & 72\% (23)\\
        \hline
        n5\_d3 & 19$\rightarrow$20 & 1,024 & 1 & yes & yes & < 1s & 38\% (12)\\
        \hline
        n5\_d4 & 25$\rightarrow$26 & 40,894,464 & 1 & yes & yes & < 1s & 19\% (6)\\
        \hline
    \end{tabular}
    \caption{Stagnation \& Functional Collapse (RQ3)}
    \label{tab:stagnation}
\end{table}
Table \ref{tab:stagnation} shows the results of detecting stagnation in \textsc{ApproxMC} and verifying functional equivalence across different experimental settings. Column 1 shows the tree topology. For each experiment, we identify the step in Column 2, showing at which stage that the approximate count of remaining hypotheses stops decreasing, indicating stagnation. The before and after approximating model numbers are shown in Column 3 and 4. At these steps, the functional equivalence of the remaining hypotheses is verified using a SAT solver, confirming that the hypothesis space has collapsed to a unique solution. Column 5 and 6 indicates if a stagnation is found and if the functional equivalence is verified.

The experiments show that in all cases, the SAT check completes in less than 1 second, demonstrating the efficiency of the functional equivalence check. The count before stagnation varies depending on the problem size (e.g., 100 hypotheses for n=4, d=2 and 23,552 hypotheses for n=4, d=3, but the method consistently reduces the hypothesis space to a single solution in very few steps. This indicates that our active learning method is effective at quickly identifying the correct Boolean function, with functional collapse detected at each step of stagnation, confirming the validity of the approach. The last column shows the percentage (the number of queries) our method reduced from the all features permutation queries (e.g., 4 features which require $2^4 = 16$ queries). For all trees that stagnate during the learning process, at least 19\%/6 queries and at most 72\%/23 queries were saved. They also demonstrate our tool's effectiveness on query number optimization.

\section{Discussion}
\label{sec:dis}

In this section, we discuss theoretical and practical considerations for scaling to larger trees, and the limitations of our methods.

\paragraph{\textbf{Scalability Considerations}:} The full hypothesis space for a depth-$d$ decision tree with $n$ Boolean features has size  $n^{(2^d - 1)} \cdot 2^{2^d}$, where the first factor enumerates feature assignments for internal nodes, and the second assigns outputs to leaves. This space grows exponentially in $d$, making explicit enumeration infeasible beyond shallow depths. Our approach remains tractable on our experimental settings because the entire space is represented symbolically using CNF, avoiding explicit enumeration. \texttt{ApproxMC} provides approximate counts in time sublinear in $|\mathcal{H}|$, using hashing-based universal model counting~\cite{DBLP:conf/aaai/ChakrabortyMMV16}. Each highly selected query reduces the version space by a factor close to one half evidenced in our experiments. One might also expect each membership query to eliminate roughly half of the candidate trees, since decision trees partition the input space at each node. In reality, a query constrains only the leaves and internal nodes along its input path, leaving many other hypotheses consistent. The fraction of eliminated trees varies per query and is rarely exactly 50\%. By using approximate model counting, our method quantitatively estimates the version-space reduction for each candidate query efficiently. It is worthwhile to note that \texttt{ApproxMC} may be unable to effectively scale up if the function space is inherently hard to reduce.

\paragraph{\textbf{Limitations}:} Several limitations of our method should be acknowledged. Since the runtime of our algorithm scales with the cost of approximate model counting, which can still be nontrivial for large CNFs. We can handle as many as 10k CNF clauses with over 1k variables. Furthermore, when the internal nodes support more than binary decisions, the formula complexity would also grow significantly. There may be a better strategy to balance the optimality for each round and time. For example, a parallel algorithm can be leveraged in the query selection procedure to speed up 10x with 10 threads, which can approximate the models for different inputs in parallel. However, our focus here is to find the approximately optimal choice iteratively, and we leave the parallel aspect for future directions. Another threat is \texttt{ApproxMC}-based estimates may occasionally select suboptimal queries if estimation error skews comparison. Our method also does not provide theoretical query complexity guarantees, where no worst-bounds results are achieved. Finally, our evaluation assumes a noise-free oracle. In realistic settings with noisy or adversarial labels~\cite{DBLP:conf/colt/Cesa-BianchiGVZ10}, space elimination may become inconsistent, and the current SAT-based mechanism offers no robustness guarantees. Despite these constraints, the empirical behavior observed in Section~\ref{sec:eval} confirms that the algorithm remains practical and robust for medium-scaled decision trees.

\section{Related Work}
\label{sec:rel}

Classical exact active learning was introduced by Angluin~\cite{DBLP:journals/iandc/Angluin80, DBLP:journals/csur/AngliunS83}, who studied learning in the exact identification setting with membership queries. Decision-tree inference from queries has also been explored in theory~\cite{DBLP:journals/mor/GuptaNR17, DBLP:journals/orl/ZhuoN26}, including bounds on optimal decision tree learning and information gain heuristics. Our contribution differs in that we do not operate over an explicit enumeration of tree hypotheses; instead, we maintain a \emph{symbolic} representation in CNF and use constraints to guarantee consistency.

Model counting has a long history in formal verification and probabilistic reasoning. Exact model counters such as \textsc{sharpSAT}~\cite{DBLP:conf/sat/Thurley06} achieve high performance in practice but still remain worst-case exponential. Approximate counters such as \texttt{ApproxMC}~\cite{DBLP:conf/cp/ChakrabortyMV13} provide provable $(\varepsilon,\delta)$ guarantees using universal hashing. The idea of using approximate model counts to guide search has appeared in domains such as probabilistic inference~\cite{DBLP:journals/ai/ChaviraD08, DBLP:conf/cp/DubraySN23}, but, to our knowledge, has not previously been used to \emph{drive optimal active query selection} in formal learning frameworks.

SAT and SMT-based learning from examples has been explored in inductive program synthesis~\cite{DBLP:conf/fmcad/AlurBJMRSSSTU13, DBLP:conf/oopsla/PolozovG15}, repair~\cite{DBLP:journals/tse/GouesHSBDFW15, DBLP:conf/icse/Huang024}, and invariant generation~\cite{DBLP:journals/corr/KrishnaPW15, DBLP:conf/popl/0001NMR16}. These works typically add constraints incrementally as new counter examples learned, but they rely either on heuristic search or program abstract domain rather than numerical estimation of version space reduction. In contrast, our system quantifies the remaining hypothesis space volume after each candidate query, allowing a principled halving strategy analogous to classical active learning but with no need to enumerate the version space explicitly.

\section{Conclusion}
\label{sec:conc}

We introduced a new framework for active learning in which the hypothesis space is represented symbolically in CNF and the expected information gain of each candidate query is estimated using approximate model counting. This enables a learning strategy without enumerating the hypothesis space, even when the number of possible functions is combinatorially huge. We demonstrated this approach using various numbers of depth and features of unknown decision trees as targets, encoding them in Boolean variables. \texttt{ApproxMC} was used to estimate, for each possible query input, the approximate number of hypotheses consistent with each oracle response. The input minimizing the estimated remaining version space was chosen automatically, and incremental solving strategy enforced consistency with observed answers. We showed that when \texttt{ApproxMC} stagnates, a functional collapse check is performed to verify a functional equivalent tree is found. For future work, we plan to extend our method to more applications and discover more efficient oracle input selection strategies.

\paragraph{\textbf{Acknowledgments}} Zunchen Huang and Chenglu Jin are (partially) supported by project CiCS of the research programme Gravitation, which is (partly) financed by the Dutch Research Council (NWO) under the grant 024.006.037. We thank Chao Yin, Marten van Dijk, and Fabio Massacci for their discussions on function recovery on gate-hiding garbled circuits, providing some insights on this work.
%
%
%

\bibliographystyle{splncs04}
\bibliography{main}

\newpage
\appendix
\section*{Appendix}
\textbf{Proof for Lemma~\ref{lem}}:

\begin{proof}
    A full binary tree of depth $d$ (root at depth $0$) has exactly
    \[
    N_{\text{leaf}} = 2^{d}
    \]
    leaves and
    \[
    N_{\text{int}} = 2^{d}-1
    \]
    internal (non-leaf) nodes.  This is standard: each level $k$ ($0\le k < d$)
    contains $2^{k}$ internal nodes, and summing yields
    $\sum_{k=0}^{d-1} 2^{k} = 2^{d}-1$.
    
    A hypothesis in $\mathcal{H}_{d,n}$ is fully specified by two independent choices:
    
    \begin{enumerate}
        \item For every internal node (there are $N_{\text{int}}$ of them) choose one feature from the $n$ available features. Since selections are independent across nodes and repetition is allowed, the number of ways to assign features to all internal nodes is
        \[
        n^{\,N_{\text{int}}} = n^{\,2^{d}-1}
        \]
        
        \item For every leaf (there are $N_{\text{leaf}}$ of them) choose a Boolean output in $\{0,1\}$. The number of ways to label all leaves is
        \[
        2^{\,N_{\text{leaf}}} = 2^{\,2^{d}}.
        \]
    \end{enumerate}
    
    Because these two choices are independent (feature selections do not constrain leaf labels and vice versa), the total number of hypotheses equals the product of the two counts:
    \[
    |\mathcal{H}_{d,n}| \;=\; n^{\,N_{\text{int}}} \cdot 2^{\,N_{\text{leaf}}}
    \;=\; n^{\,2^d-1} \cdot 2^{\,2^d}.
    \]
    This completes the proof.
    \qed
\end{proof}




\end{document}